\documentclass[aps,nofootinbib,amsfonts,groupedaddress]{revtex4}
\usepackage{amsmath}
\usepackage{epsfig,epsf}
\usepackage{graphicx}
\usepackage{verbatim}
\usepackage{epstopdf}

\def\beq{\begin{equation}}
\def\eeq{\end{equation}}
\def\bea{\begin{eqnarray}}
\def\eea{\end{eqnarray}}
\def\eq#1{{Eq.~(\ref{#1})}}
\def\fig#1{{Fig.~\ref{#1}}}

\newcommand{\bas}{\bar{\alpha}_s}
\newcommand{\as}{\alpha_s}
\newcommand{\Lb}{\left(}
\newcommand{\Rb}{\right)}
\newcommand{\nn}{\nonumber}

\newcommand{\h}{\frac{1}{2}}

\def\pom{{I\!\!P}}
\begin{document}


\voffset1.5cm
\title{The Ridge  from the BFKL evolution and beyond}
\author{Eugene Levin$^{1,2,3}$ and Amir H. Rezaeian$^1$}
\affiliation{
$^1$ Departamento de F\'\i sica, Universidad T\'ecnica
Federico Santa Mar\'\i a, Avda. Espa\~na 1680,
Casilla 110-V, Valparaiso, Chile \\
$^2$Centro Cientifico-Tecnol$\acute{o}$gico de Valpara\'\i so,
Casilla 110-V,  Valparaiso, Chile\\
$^3$ Department of Particle Physics, Tel Aviv University, Tel Aviv 69978, Israel}

\date{\today}
\begin{abstract}
We show that the long-range rapidity correlations between the produced
charged-hadron pairs from two BFKL parton showers generate
considerable azimuthal angle correlations. These correlations have no
$1/N_c$ suppression. The effect of gluon saturation on these
correlations are discussed and we show that it is important. We show
that a pronounced ridge-like structure emerges by going from the BFKL to the
saturation region. We show that the ridge structure at high-energy
proton-proton and nucleus-nucleus collisions has the same origin and
its main feature can be understood due to initial-state effects.
Although the effects of final-state interactions in the latter case can be non-negligible. 
\end{abstract}
\maketitle


\section{Introduction}

  The main objective of this paper is to understand the long range
  rapidity correlations of charged-particle pairs in the azimuthal
  angle separation between the two particles around the near side
  $\Delta\varphi\approx 0$, the so-called ridge which has been recently
  observed at the LHC in $\sqrt{s}=2.76$ TeV Pb+Pb collisions
  \cite{CMSnew} and also in $\sqrt{s}=7$ TeV proton-proton ($pp$)
  collisions \cite{CMSLRC}. The CMS collaboration \cite{CMSLRC}
  recently reported that the ridge type structure exists in $pp$
  collisions at $\sqrt{s}=7$ TeV for high multiplicity $N\ge 90$ event
  selections. The origin of the ridge in $pp$ collisions at the LHC is
  not still well understood and it has been a subject of growing
  interest, see for example Refs.~\cite{NEWCORPP,ppm,rest}.  The ridge
  was previously seen at RHIC in central Cu+Cu collisions at
  $\sqrt{s}=62.4$ GeV and in Au+Au collisions at $\sqrt{s}=200$ GeV
  \cite{RHICCOR}. The description of nucleus-nucleus ($AA$) collisions
  is generally more complicated compared to the case of $pp$
  collisions. However, given the relative similarity of the observed
  ridge structure in both $pp$ and $AA$ collisions in terms of
  multiplicity, transverse momenta and rapidity separations of pairs,
  it is natural to ask whether the ridge phenomenon has a unique
  origin and can be understood only by initial-state effects. We
  recall that the highest multiplicity events per unit rapidity in
  $pp$ collisions at $\sqrt{s}=7$ TeV is compatible to the one in
  central Cu+Cu collisions at RHIC.

 In high density QCD, we expect large rapidity correlations for
 produced hadron pairs with the value of their transverse momenta
 about the gluon saturation scale $Q_s$
 \cite{THEOREXP,NEWCOR1,NEWCORPP}, see also Ref.~\cite{more-r}.  At
 first sight, these correlations should be small at fixed impact
 parameters. It has been argued \cite{NEWCOR1} that in the
 color-glass-condensate (CGC) approach \cite{MV} there is a source of
 the long range rapidity correlations which transforms into the
 azimuthal angle correlations due to the collective flow in the final
 state. In Ref.~\cite{NEWCORPP} it is argued that such mechanism can
 qualitatively explain the azimuthal angle correlations in
 proton-proton collisions without a significant flow effect. The issue
 of the importance of final-state and collective flow effects in the
 observed ridge structure in $pp$ collisions \cite{CMSLRC} is still debatable
 \cite{ppm}, see also Refs.~\cite{dav,mev2}.

In this paper we will introduce a new source of long-range azimuthal
correlations for the produced charged hadron pairs. We show that the
intrinsic long-range rapidity correlations between the produced hadron
pairs from two parton showers generate considerable azimuthal angle
correlations which do not depend on the interaction in the final
state, and because of this, these correlations have the same origin
both in $pp$ and $AA$ interactions at high energy. These correlations
have no $1/N_c$ suppression as one considered in
Refs.~\cite{NEWCOR1,NEWCORPP}. Recently Kovner and Lublinsky in a very
nice paper \cite{KOVLUCOR} put forward a general discussion toward
understanding the ridge. We have an additional goal here, like
Ref.~\cite{KOVLUCOR} we shall try in this paper to understand the
general feature of the ridge based on very general grounds and will
show that the main features of these correlations both in rapidity and
emission angle can be simply understood within the BFKL Pomeron
calculus \cite{BFKL,GLR,MUQI,BRN,BART,LMP}. The extension beyond this
framework inside the saturation regime will be also discussed.

The paper is organized as follows: In sec II, we introduce our
mechanism for the azimuthal correlations and illustrate the main idea
within the perturbative framework. In Sec. III, we consider double
inclusive gluon production and its correlations within the BFKL
Pomeron approach. We show that the azimuthal correlations between
produced hadron pairs from two BFKL parton showers have long-range
nature and will survive the BFKL leading log-s re-summation. In
Sec. IV, we provide estimates of azimuthal correlations in both $pp$
and $AA$ collisions in the BFKL and the saturation regions. As a
conclusion, in Sec. V we highlight our main results.

\section{The azimuthal correlations: the origin }
\begin{figure}[t]
\begin{center}
       \includegraphics[width=12cm] {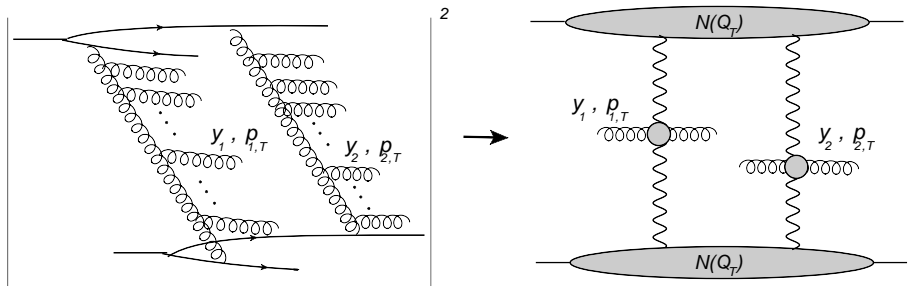}
\end{center}
\caption{Mueller diagrams for two parton showers production. The wave lines denote the BFKL Pomerons. 
This is the typical diagram which gives an angular collimation about $\Delta \varphi\approx 0$.  } 
\label{2psh}
\end{figure}

\begin{figure}[t]
\begin{center}
        \includegraphics[width=12cm,height=2cm] {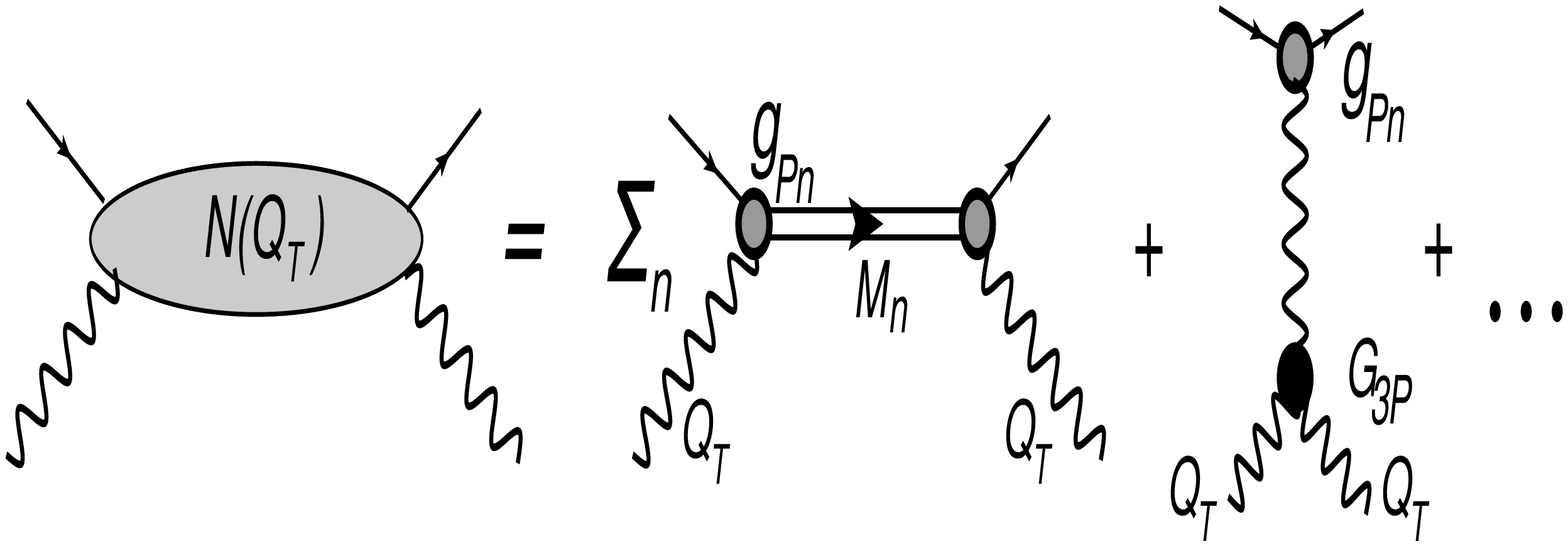}
\end{center}
\caption{Diagrams representing the Pomeron-hadron scattering amplitude $N_{\pom h}\Lb Q_T\Rb$ as a sum of resonance contributions, 
triple-Pomeron diagram with the vertex denoted by $G_{3 \pom}$ and
etc, see the text for the details. The wave lines denote the Pomeron,
while lines represent hadrons.}
\label{Npom}
\end{figure}

In this section, we show that the long range rapidity correlations in 
azimuthal angle separation between the hadron pairs can be simply understood in the perturbative QCD approach. 
In the parton-like language, the Mueller diagram \cite{MUDIA} shown in
\fig{2psh} (right panel) describes the emission of two particles
(partons) from two parton showers. One can write the contribution of
this diagram to the cross-section of double inclusive gluon production
in the following generic form, 
\beq
\label{BFKLEM1} 
\frac{d \sigma}{ d y_1 d^2 \vec p_1\,d y_2 d^2
\vec p_2}\,\,=\,\,\h\,\int d^2 \vec Q_T \,N_{\pom h}^2(Q_T^2)\, \frac{d \sigma}{d y_1d^2
\vec p_1}\Lb \vec{Q}_T\Rb
\,\frac{d \sigma}{d y_2d^2 \vec p_2}\Lb -\vec{Q}_T\Rb,
\eeq 
where $N_{\pom h}$ is the scattering amplitudes for Pomeron (ladder)-hadron
productions along which transverse momentum $\vec Q_T$ is transferred and $
d\sigma/d y_{i}d^2 \vec p_{iT}$ denotes the corresponding cross-section of the gluon
production with rapidity $y_i$ and $\vec p_{iT}$ in each of the BFKL Pomeron
ladders. This factorization is based on the leading Log-s approximation
ignoring enhanced Pomeron diagrams.
\eq{BFKLEM1} can be motivated \cite{COL,SOFT,LRCORLR,LRCORKT,LRCORORSAY}
using three main ingredients: Gribov Reggeon \cite{COL,SOFT} and
Pomeron \cite{BFKL,GLR,MUQI,BRN,BART} calculus, AGK cutting rules
\cite{AGK} and Mueller generalized optical theorem \cite{MUDIA}. In
the Pomeron calculus the amplitude $N_{\pom h}$ is a new ingredient which can be written in the following form,
\beq \label{EQN}
N_{\pom h}\Lb Q_T\Rb \,\,=\,\,\sum^{M_{max}}_{n =1}\,g^2_{\pom n}\Lb Q_T\Rb\,\,+\,\,\int^{\infty}_{M_{max}} \frac{d M^2}{M^2}\,g_{\pom p}\Lb Q_T=0\Rb
\,G_{3\pom }\Lb Q_T\Rb\,\,\Lb M^2/s_0\Rb^{-\Delta_\pom}+\dots,
\eeq
where $n$ denotes the number of produced state with mass $M_n$ in the
diffractive dissociation with $M_{max}$ as its maximum value
(about $2$ GeV) by which one can still
express $N_{\pom}$ as a sum of resonances, $g_{\pom n}$ denotes the
vertex of the Pomeron with this state ($g_{\pom n}=g_{\pom p}$ for
$n=1$) and $G_{3\pom}$ denotes triple Pomeron vertex. The first term
in \eq{EQN} describe the contribution of the state with finite mass
and this sum can be approximated by the sum of produced
resonances. The second term is responsible for high mass contribution
and can be described by the Pomeron contribution which leads to the
factor $\Lb M^2/s_0\Rb^{-
\Delta_\pom}$ where $\Delta_\pom$ is the Pomeron intercept and $s_0$ is the
energy scale ($s_0 \approx 1\,\text{GeV}$) \cite{LRCORORSAY}, see
\fig{Npom}.  In the framework of the high energy Pomeron phenomenology
it turns out that $Q_T$ dependence of the resonance contribution
is much steeper than the one in the triple Pomeron term. In the BFKL Pomeron
calculus this fact has a natural explanation: the resonance
contributions are determined by the non-perturbative soft scale which
is about 1 fm, while the triple BFKL Pomeron vertex has a natural scale
of the order of the saturation scale which increases with energy.  It
should be stressed that $N_{\pom h}$ has a very simple physical meaning, namely
$N_{\pom h}^2$ is the probability to produce two parton-showers in
hadron-hadron collisions.

\begin{figure}[t]
\begin{tabular}{c c c}
       \includegraphics[height=4.5cm, width=10cm] {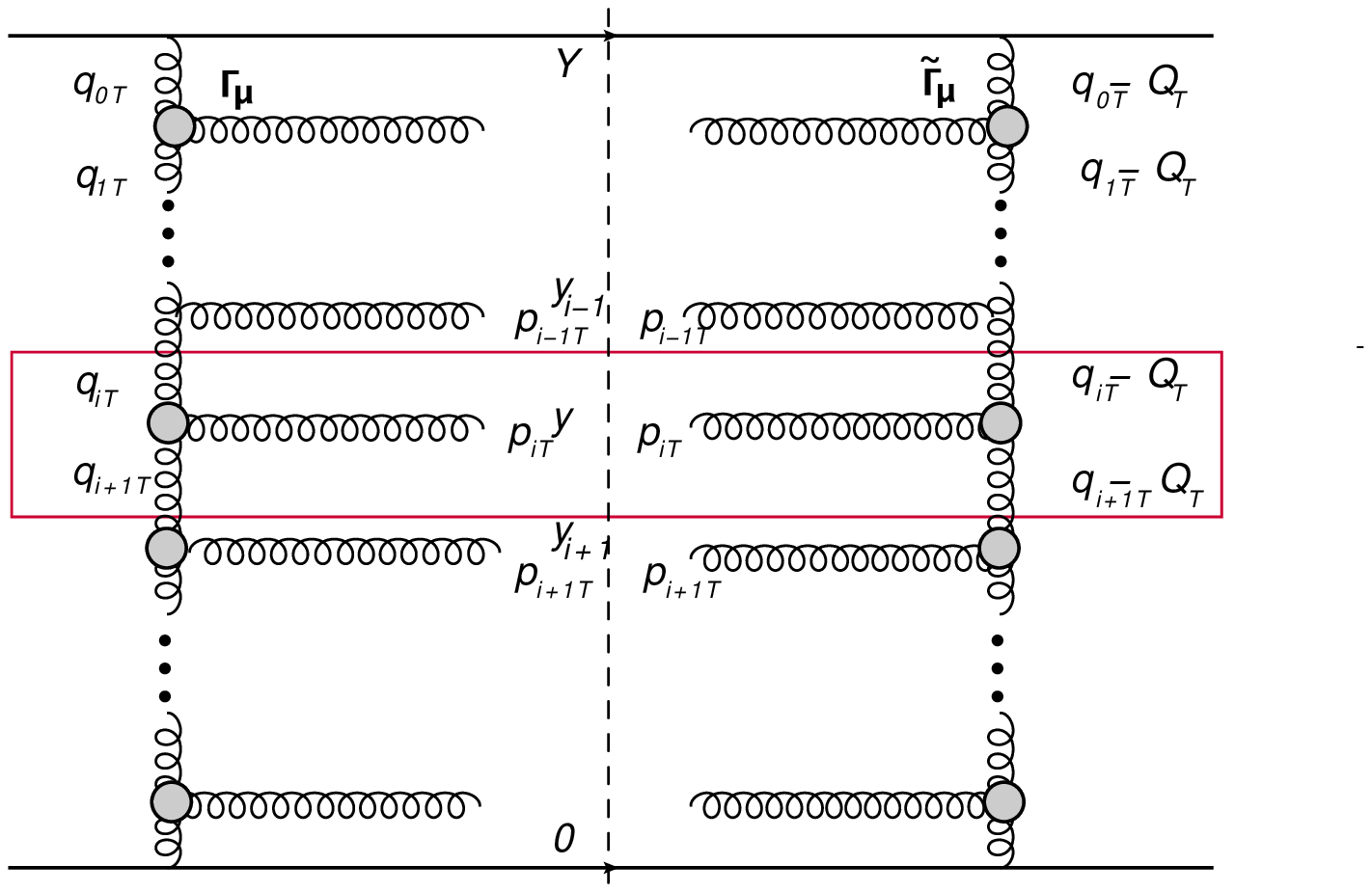} && \includegraphics[height=4cm, width=6cm] {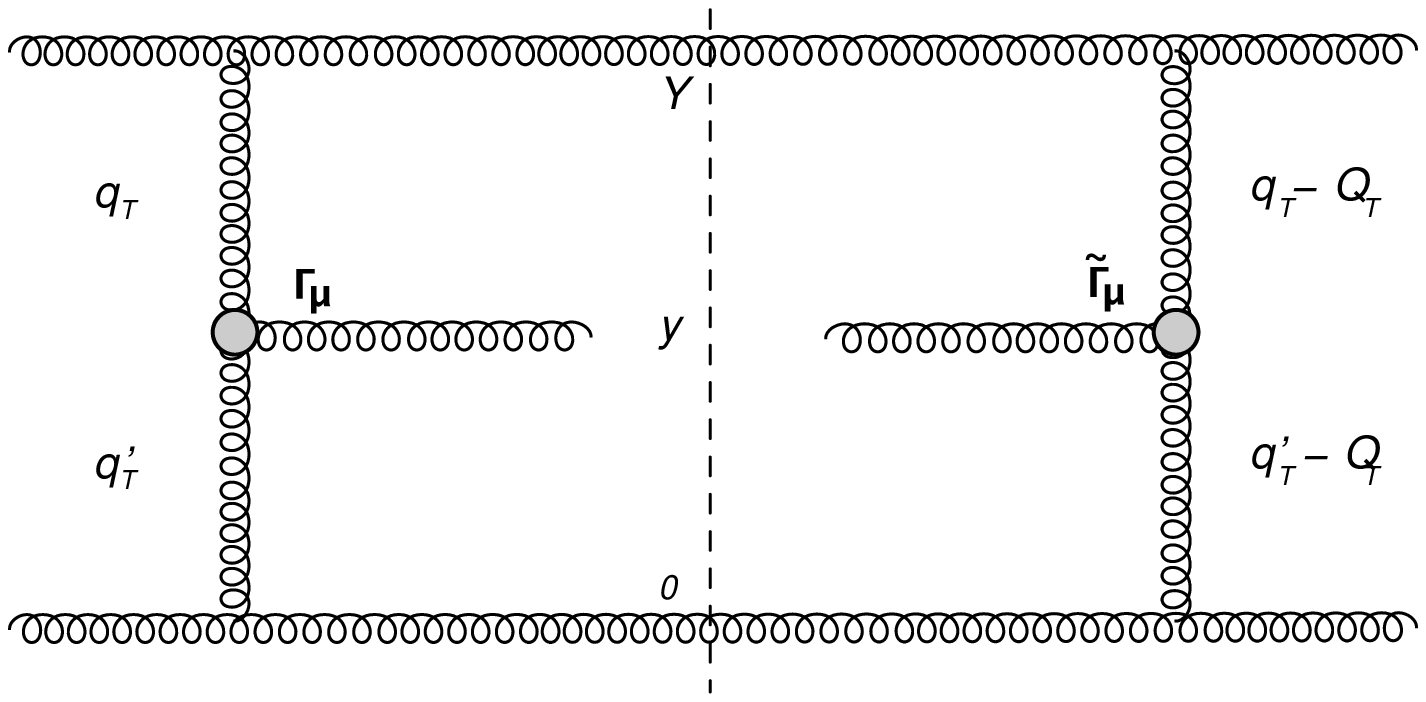}\\
\fig{pomQt}-b & &\fig{pomQt}-a\\
\end{tabular}
\caption{The ladder-type diagram that describes the production of gluon with transverse momentum $p_{iT}$
in the Born approximation (\fig{pomQt}-a) and BFKL Pomeron
(\fig{pomQt}-b). The blobs represent Lipatov vertices and asterisks in
left denote reggeized gluons. The produced gluon in the $i$th rung is shown within a box. }
\label{pomQt}
\end{figure}

At first sight, one may expect that \fig{2psh} describes two
independent parton showers, and therefore there should not be any
correlation between two produced gluons from these two parton showers.
However, angular correlations stem from the $\vec Q_T$ integration in
\eq{BFKLEM1}. Due to this integration the contribution of diagram
in \fig{2psh} is not equal to the product of two single inclusive
cross-sections leading to nonzero two particle correlation $\mathcal R \neq 0$. In order to illustrate this
simple fact, let us for the sake of argument assume that the gluon
production cross-section in one parton shower is proportional to
$\vec{Q}_T \cdot \vec{p}_{i,T}$, or in other words,
\beq 
\frac{d \sigma}{d y_id^2 \vec p_i}(Q_T)\propto \vec{Q}_T \cdot \vec{p}_{i,T}\frac{d \tilde{ \sigma}}{d^2 y_id^2 \vec p_i}.
\eeq
In this case, \eq{BFKLEM1} simply becomes
\bea \label{BFKLEM11}
 \frac{d \sigma}{ d y_1 d^2 \vec p_{1,T}\,d y_2 d^2 \vec p_{2,T}} \,\,&\propto&\,\, \int d^2 \vec Q_T \,N_{\pom h}^2(Q^2_T)\, \frac{d \tilde{ \sigma}}{d y_1 d^2 \vec p_{1,T}}\Lb Q^2_T\Rb
\,\frac{d \tilde{\sigma}}{d y_2 d^2 \vec p_{2,T}}\Lb Q^2_T\Rb \Lb\vec{Q}_T \cdot \vec{p}_{1,T}\Rb\,\Lb\vec{Q}_T \cdot \vec{p}_{2,T}\Rb \nn,\\
 & =& -\,\vec{p}_{1,T}\cdot\vec{p}_{2,T}\,(\pi/2) \int d Q^2_T \,N_{\pom h}^2(Q^2_T)\, \frac{d \tilde{ \sigma}}{d y_1d^2 \vec p_{1,T}}\Lb Q^2_T\Rb
\,\frac{d\tilde{ \sigma}}{d y_2d^2 \vec p_{2,T}}\Lb Q^2_T\Rb. 
\eea
The above equation explicitly shows an angular correlation between two
produced gluons in two parton showers.  Having this equation in mind,
in the next section we will explicitly show that the
vertex emission of gluon from the BFKL Pomeron with $Q_T \neq 0$ (see \fig{pomQt}) have a structure similar to Eq.~(\ref{BFKLEM11}). 

 For simplicity and clarity of the presentation, let us first work in
 the Born approximation, see \fig{pomQt}-a. In this approximation up
 to $\alpha^3$ strong-coupling corrections, the inclusive singlet gluon
 production at very high energy, assuming that all components of the
 exchanged momentum are much smaller than the projectile and target
 momentum (for $s>> \mid t\mid$), is given by
\bea
\frac{d^2 \sigma}{d y\, d^2 \vec p_T}\,\,&=&\,\,\frac{2\alpha^3C_F}{\pi^2}\int d^2 \vec q_T\,\frac{\Gamma_\mu\Lb \vec q_T, \vec{q'}_T\Rb\, \tilde{\Gamma}^\mu\Lb -(\vec q-\vec Q)_T, -(\vec{q'}- \vec{Q})_T\Rb}{ q^2_T \,( \vec Q - \vec q)^2_T\,q'^2_T\,(\vec Q - \vec{q'})^2_T}\,\,\label{BFKLEM2}\,,
\eea
where $C_F=(N^2_c-1)/2N_c$ is the $SU(N_c)$ Casimir operator in the
fundamental representation with the number of color equals $N_c$. We
used a notation $\vec{p}_T \,=\,\vec{q}_T - \vec{q'}_T$. The effective
vertex $\Gamma_\mu$ and $\tilde{\Gamma}_\mu$ for the emission of
gluons (see
\fig{pomQt}-a) are related to the Lipatov vertex
$\Gamma_{\mu\nu}^\rho$ \cite{BFKL,GLR} in the following way,
\beq \label{G}
\tilde{\Gamma}^\rho\Lb \vec q_T, \vec{q'}_T\Rb\,\,=\frac{2}{s}p_{1\mu}p_{2\nu}\Gamma_{\mu\nu}^\rho\Lb \vec q_T, \vec{q'}_T\Rb,
\eeq
where $p_{1}$ and $p_{2}$ represent the momenta of the incoming
projectile and target gluon, and the center of mass energy is
$s=2\vec p_1.\vec p_2$. The product of the two vertices appeared in \eq{BFKLEM2} can be simplified to, 
\bea \label{GG}
K\Big(\vec Q_T; \vec q_T, \vec{q'}_T\Big)\,\,&\equiv&\,\,\frac{1}{2}\Gamma_\mu\Lb \vec{q'}_T, \vec{q}_T\Rb\, \tilde{\Gamma}^\mu\Lb -(\vec q-\vec Q)_T, -(\vec{q'}- \vec{Q})_T \Rb\,\,,\nonumber\\
&=&\,\frac{1}{p^2_T}\,\Big(  q'^2_T( \vec{Q} - \vec{q})^2_T \,\,+ \,\, q^2_T\,(\vec{Q} - \vec{q'})^2_T \,\,-\,\,p^2_T\,Q^2_T\Big).
\eea
 Substituting the above
expression into the cross-section \eq{BFKLEM2}, one immediately obtains 
\bea
\frac{d^2 \sigma}{d y\, d^2 \vec p_T}
  & \propto& \,\,\alpha^3\, \int \frac{ d^2 \vec q_T}{p^2_T}\,\Big( \frac{1}{q'^2_T\,(\vec Q - \vec q)^2_T}\,\,+\,\,\frac{1}{q^2_T\,(\vec Q - \vec{q'})^2_T}\,\,-\,\,\frac{Q^2_T p^2_T}{q^2_T\,q'^2_T\,(\vec Q - \vec q)^2_T\,( \vec Q - \vec{q'})^2_T}\Big),\label{GG0}\\
&\xrightarrow{ p_T \,\ll \, q_T; \,Q_T\, \ll\, q_T} \,\,\,\,\,& \alpha^3\,\int \,\frac{d^2 \,\vec q_T}{ p^2_T\,q^4_T}\,\,\Big\{2\,\, +\,\,4\,\frac{ \vec{Q}_T \cdot
\vec{p}_T}{ q^2_T}\,\,+\,\,32 \frac{\Lb \vec{p}_T \cdot \vec{Q}_T\Rb^2}{q^4_T}\Big\}\,,\label{BFKLEM3}\\
&\xrightarrow{ p_T \,\gg \, q_T; \,Q_T\, \ll\, q_T} \,\,\,\,\,& \alpha^3\,\int \frac{d^2 \vec q_T}{q^2_T\,p^4_T}\,\Big\{ 2\,\,+\,\,2\frac{\vec{Q}_T\cdot \vec{p}_T}{p^2_T}\,\,+\,\,4\,\frac{\Lb \vec{p}_T\cdot \vec{Q}_T\Rb^2}{p^4_T}\Big\}.\label{BFKLEM4}
\eea
Notice that in the Born approximation we do not consider the kinematic
region $q_T\, \ll \,Q_T$ since we will show later that this region is
not important for the azimuthal correlations from the BFKL Pomeron.
Moreover, we should stress that the expansion here are only for the
purpose of illustration to trace back the origin of the azimuthal
angle correlations in our approach while for the practical estimates,
one has to perform the integrals without resorting to any
approximation.

First notice that \eq{BFKLEM2} is symmetric\footnote{We thank our referee
for drawing our attention to this point.} under $\vec{q}_T \to
\vec{q'}_T$ and $ \vec{p}_T\to -\vec{p}_T$. In the expansion given in Eqs.~(\ref{BFKLEM3},\ref{BFKLEM4}), we changed
the variable to $\vec{q'}_T \,=\,\vec{q}_T\,-\,\vec{p}_T$. Changing the variable in \eq{GG0}
to $\vec{q}_T \,=\,\vec{q'}_T\,+\,\vec{p}_T$ and then in the same fashion expanding we get
the same expression as in the above equations but the second term in
Eqs.~(\ref{BFKLEM3},\ref{BFKLEM4}) will be with the opposite
sign. Actually, these two expansions correspond to different regions
of integrand in \eq{GG0}. Summing these two
contributions\footnote{These two expansions can be also envisaged as
two different processes: in Eqs.~(\ref{BFKLEM3},\ref{BFKLEM4}) the
transverse momentum of produced gluon is compensated by the gluon with
the value of the rapidity smaller than the rapidity of the produced
gluon with the transverse momentum $p_T$ (gluon with rapidity 0 in
\fig{pomQt}-a), while expansion in $q'_T$ we consider the process
where $p_T$ is balanced by the gluon with the rapidity larger than the
rapidity of the produced gluon with the transverse momentum $p_T$ (gluon with rapidity $Y$ in
\fig{pomQt}-a).}  we obtain the following form for the double
inclusive cross-section from
\eq{BFKLEM1} in the case of $\vec p_T \,\ll \, \vec q_T, \vec{q'}_T;
\,\vec Q_T\,
\ll\, \vec{q'}_T, \vec q_T$,
\bea \label{BFKLEM5}
&& \frac{d \sigma}{ d y_1\,d y_2\,d^2 \vec p_{1,T}\,d^2 \vec p_{2,T}}\,\,=\,\,\int d^2 \vec Q_T N_{\pom h}^2\Lb Q_T\Rb\,
\frac{d \sigma}{ d y_1\,d^2 \vec p_{1,T}}\Lb Q_T = 0\Rb\,\frac{d \sigma}{ d y_2\,d^2 \vec p_{2,T}}\Lb Q_T = 0\Rb\,\,\nn\\
&&+ \,\,32~ p^2_{1,T}p^2_{2,T}\Lb 2 \,+\,\cos\left(2 \Delta\varphi\right)\Rb\int d^2 \vec Q_T\,Q^4_T\, N_{\pom h}^2\Lb Q_T\Rb\,\frac{d \tilde{\sigma}}{ d y_1\,d^2 \vec p_{1,T}}\Lb Q_T = 0\Rb\,\frac{d \tilde{\sigma}}{ d y_2\,d^2 \vec p_{2,T}}\Lb Q_T = 0\Rb,\nonumber\\
\eea
where  $\Delta\varphi$ denotes the angle between $\vec{p}_{1,T}$ and $\vec{p}_{2,T}$ and we defined
\beq \label{BFKLEM6}
\frac{d \sigma}{ d y d^2 \vec p_T}\,\,=\,\,4 \frac{ 2 \as}{C_F}\,\frac{1}{p_T^2}\,\int\,d^2 \vec q_T \phi\Lb \vec q_T, -\vec q_T\Rb\,\phi\Lb\vec{q}_T - \vec{p}_T, \vec{p}_T - \vec{q}_T\Rb,
\eeq
and
\beq  \label{BFKLEM7}
\frac{d \tilde{\sigma}}{ d y d^2 \vec p_T}\,\,=\,\,4 \frac{ 2 \as}{C_F}\,\frac{1}{p_T^2}\,\int\,\frac{d^2 \vec q_T}{q^4_T}\phi\Lb \vec q_T, -\vec q_T \Rb\,\phi\Lb\vec{q}_T - \vec{p}_T, \vec{p}_T - \vec{q}_T\Rb =\langle \frac{1}{q_T^4}\rangle \frac{d \sigma}{ d y d^2 \vec p_T}.\
\eeq
In the above, we used the following notation,
\beq \label{ava-q}
\langle \frac{1}{q_T^4}\rangle =
\frac{\int\,\frac{d^2 \vec q_T}{q^4_T}\phi\Lb \vec{q}_T ,- \vec{q}_T\Rb\,\phi\Lb\vec{q}_T - \vec{p}_T,\vec{p}_T - \vec{q}_T\Rb}
{\int\, d^2 \vec q_T\phi\Lb\vec q_T,-\vec q_T \Rb\,\phi\Lb\vec{q}_T - \vec{p}_T,\vec{p}_T - \vec{q}_T\Rb},
\eeq
where $\phi$ denotes the unintegrated gluon density of the projectiles \cite{yura} for $Q_T=0$, 
\beq \label{BFKLEM8}
\phi\Lb \vec q_T,-\vec q_T \Rb\,\,=\,\,\frac{\as\,C_F}{\pi}\,\frac{1}{q^2_T}.
\eeq
Notice that at $Q=0$ the inclusive cross-section given in \eq{BFKLEM2} is identical to \eq{BFKLEM6}.

This simple example indicates that we have a natural mechanism for the
azimuthal correlations in the framework of perturbative QCD which does
not depend on the final state interactions and leads to the
correlations inside of initial wave-function of the incoming hadrons. In the next section,
we will show that this azimuthal correlation has long-range nature and will
survive the BFKL leading log-s resummation.

\section{Long-range azimuthal correlations for two BFKL parton showers}
The generalization of the Born approximation to the case of
gluon emissions from the BFKL Pomeron cannot be simply obtained via \eq{BFKLEM5} by 
replacing the unintegrated gluon density $\phi$ to the one
obtained from the BFKL equation. Indeed, the unintegrated gluon density $\phi$ depends also on $\vec Q_T$ and we have to be very careful
with putting $Q_T = 0$.  The inclusive gluon product can be generally written as 
\beq  \label{2BFKLSH1}
\frac{d \sigma\Lb Q_T\Rb}{ d y d^2 \vec p_T}\,\,=\,\,4 \frac{ 2 \as}{C_F}\,\int d^2 \vec q_T\,K\Big(\vec Q_T; \vec q_T, \vec{q'}_T\Big)\,\frac{1}{q'^2_T\,(\vec{Q} - \vec q)^2_T}\,\phi\Lb Y - y , \vec q_T, \vec{Q}_T  - \vec{q}_T\Rb\,\phi\Lb y, \vec{q}_T - \vec{p}_T, \vec{Q}_T - \vec{q}_T + \vec{p}_T \Rb,
\eeq
where $K\Big(\vec Q_T; \vec q_T, \vec{q'}_T\Big)$ is the BFKL kernel given in
\eq{GG} and we defined $\vec {q'}_T=\vec q_T-\vec p_T$. In the above, the variable $Y = \ln(s/m^2)$ denotes the total rapidity in the lab frame where $m$ is the nucleon mass and $y$ and $\vec p_T$ are the transverse momentum and rapidity of the produced gluon, respectively. 
Notice that at $Q_T=0$, the above expression has the same functional
form as the $k_T$ factorization \cite{yura}. The only dependence on $\vec p_T$ comes
from the term $\phi\Lb\vec{q}_T -
\vec{p}_T, \vec{Q}_T - \vec{q}_T + \vec{p}_T \Rb$ for which we have
the color-singlet BFKL equation \cite{BFKL,LR2REV}:
\bea \label{2BFKLSH2}
 \phi\Lb y, \vec{q'}_T, \vec{Q}_T - \vec{q}'_T \Rb \,\,&=&\,\,\frac{\bas}{\pi}\,
\int^y d y_{i+1}\Big\{\int d^2 \vec{q''}\,
K\Big(\vec Q_T; \vec{q'}_T, \vec{q''}_T\Big)\,\frac{1}{q''^2_T\,(\vec Q - \vec{q'})^2_T}\,\phi\Lb y_{i+1}, \vec{q''}_T, \vec{Q}_T  - \vec{q''}_T\Rb\, \nn\\
  &-& \,\Big( \frac{ q'^2_T}{ ( q'')_T^2\,( \vec{q'} - \vec{q''})^2_T}\,\,+\,\,\frac{ (\vec{Q} - \vec{q'})^2_T}{ (q'')_T^2\,(  \vec Q - \vec{q'}- \vec{q''})^2_T}\Big)\,\phi\Lb y_{i+1}, \vec{q'}_T, \vec{Q}_T - \vec{q}'_T \Rb \Big\},
\eea
where we defined $\bas=\alpha N_c/\pi$. We first substitute $
\phi\Lb y, \vec{q'}_T, \vec{Q}_T - \vec{q}'_T \Rb $ given in \eq{2BFKLSH2}
into \eq{2BFKLSH1} and expand the kernels of both equations up to the terms
of the order of $Q^2_T$. Then we again use \eq{2BFKLSH2} but at $Q_T=0$ and
collect all terms into $\phi\Lb y, \vec{q'}_T, \vec{q}'_T \Rb$. Therefore, we obtain the following equation,
\bea  \label{2BFKLSH3}
\frac{d \tilde{\sigma}\Lb Q_T\Rb}{ d y d^2 \vec p_T}\,\,&=&\,\,4 \frac{ \pi  \as}{C_F}\,\int d q^2_T\,K\Big(0; \vec q_T, \vec{q'}_T\Big) \,\frac{1}{q'^2_T q^2_T}\,\phi\Lb Y, \vec q_T, -\vec q_T\Rb\,\phi\Lb y, \vec{q'}_T, -\vec{q'}_T \Rb\Big\{ 1+\frac{\vec{p}_T \cdot \vec{Q}_T}{q'^2_T} + 2\frac{\Lb \vec{p}_T \cdot \vec{Q}_T\Rb^2}{ q'^4_T}\,\nn\\
& +&\dots +\mbox{ terms of the order of $Q_T$ that do not lead to azimuthal angle correlations} \Big\}.
\eea
In order to understand better if the above approximation can be
justified, let us examine the ladder summations which leads
to the BFKL equation. At leading log-s approximation, the imaginary
amplitude $\mathcal{A}$ of the quark-quark elastic scattering with
exchange of a color-singlet gluon ladder whose vertical lines are
reggeized gluons \cite{BFKL,GLR,LR2REV} can be written as
\beq \label{bk-f0}
\text{Im}\mathcal{A}\equiv \sum_n \mathcal{A}\Lb 2 \to n\Rb\bigotimes\,\mathcal{A}^*\Lb 2 \to n\Rb=s^2C_Fg_s^4\sum_n \int \prod_{i=0} \frac{g_s^2K\Big(\vec Q_T, \vec q_{i,T} ,\vec q_{i+1,T}\Big)}
{\vec{q}_{i+1,T}^2(\vec{q}_{i+1,T}-\vec Q_{T})^2}\left(\frac{\beta_i}{\beta_{i+1}}\right)^{\epsilon_G\Lb \vec q_{i+1,T}\Rb +\epsilon_G\Lb \vec q_{i+1,T}-\vec Q_T\Rb }, 
\eeq
where $K\Big(\vec Q_T, \vec q ,\vec{q'}\Big)$ is again the BFKL kernel given in
\eq{2BFKLSH2}. The right-hand side of the above equation shows that the BFKL Pomeron can be written as 
a sum of production cross-sections as it follows from the optical
theorem. The symbol $\bigotimes$ denotes the integrations over $n+2$-body
phase space and the parameters $\beta_i$ (with $\beta_0=1$) is the
standard Sudakov variables for the momentum of the $t$-channel gluons
which obeys strong ordering of the longitudinal momenta  \cite{BFKL,GLR,LR2REV}. The
expression in \eq{bk-f0} takes into account the reggeization of gluons
in $t$-channel that means that the spin of the gluon is not equal to
$1$ as in perturbative calculations but it is given by the reggeized
gluon trajectory
\beq \label{GR}
\alpha_G\Lb \vec q_{i,T}\Rb\,\,=1+\epsilon_G(\vec q_{i,T})=\,\, 1 \,\,+\,\,\frac{\bas}{\pi}\int \frac{d^2 \vec{q'}_T q^2_{i,T}}{q'^2_T\,( \vec{q}_{i,T} - \vec{q'}_T)^2}.
\eeq
We recall that the produced gluon in the $i$th
rung ladder is on-shell with $\vec{p}_{i,T}=\vec{q}_{i+1,T}-\vec{q}_{i,T}$. Then, in
order to find $\vec Q_T$ and $\vec p_{i,T}$ correlations one needs only to keep
$Q_T\neq 0$ in the $i$th rung of the ladder (see
\fig{pomQt}-b ) and to put $Q_T=0$ in all other rungs. The contribution 
of this particular
sell to the amplitude has the following structure
\bea \label{2BFKLSH32}
&&\frac{K\Big(\vec Q_T, \vec q_{i,T} ,\vec q_{i+1,T}\Big)}{(\vec{q}_{i+1,T}-\vec{p}_{i})^2(\vec{q}_{i+1,T}-\vec{p}_{i}-\vec Q_{T})^2
q_{i+1,T}^2(\vec{q}_{i+1,T}-\vec Q_{T})^2}\nonumber\\
&\times&
\left(\frac{\beta_i}{\beta_{i+1}}\right)^{\epsilon_G\Lb \vec q_{i+1,T}\Rb +\epsilon_G\Lb \vec q_{i+1,T}-\vec Q_T\Rb }
\left(\frac{\beta_{i-1}}{\beta_{i}}\right)^{\epsilon_G\Lb \vec q_{i+1,T}-\vec{p}_{i}\Rb +\epsilon_G\Lb \vec q_{i+1,T}-\vec{p}_{i}-\vec Q_T\Rb }.\
\eea
Although the above equation includes the virtual radiative
corrections, but has a very similar structure to the case of the Born
approximation given in \eq{BFKLEM2} and consequently in the same fashion discussed in the previous section, it also gives rise
to the azimuthal correlations. Therefore, in order to extract the
correlations between two produced gluons, it is sufficient to use \eq{2BFKLSH2} in which we can put
$Q_T=0$ in $\phi\Lb \vec{q''}_T, \vec{Q}_T - \vec{q''}_T\Rb$ and $\phi\Lb
q'_T,
\vec{Q}_T - \vec{q'}_T\Rb$. Using \eq{2BFKLSH3} and adding the contribution of the integration region in $q_T$ where 
$|\vec{q}_T -\vec{p}_T|\,\gg\,|\vec p_T|$, we obtain from \eq{BFKLEM1},  
\bea 
\frac{d \sigma}{ d y_1\,d y_2\,d^2 \vec p_{1,T}\,d^2 \vec p_{2,T}} \,\,&=&
\,\, \pi \int d Q^2_T \,N_{\pom h}^2(Q^2_T)\, \frac{d \sigma}{d y_1\,d^2 p_{1,T}}\Lb Q_T = 0\Rb
\,\frac{d \sigma}{d y_2\,d^2 p_{2,T}}\Lb Q_T = 0\Rb \,\nn \\
 & & \Big\{ 1 \,\, +\, 
\frac{1}{2}\,p^2_{1,T}\,p^2_{2,T} \,Q^4_T\,\langle \frac{1}{q^4}\rangle^2\,\Lb 2\,+\,\cos\left(2 \Delta\varphi\right) \Rb \Big\},\label{2BFKLSH4}\\
 &=& {\cal N}\,\Big( 1\,\,+\, \,\frac{1}{2}\,p^2_{1,T}\,p^2_{2,T} \,\,\langle \langle Q^4_T \rangle\rangle\,\langle\frac{1}{ q^4}\rangle^2\Lb 2\,+\,\cos\left(2 \Delta\varphi\right)\Rb \Big), \label{2BFKLSH5}
\eea
 where  $\Delta\varphi$ is the angle between $\vec{p}_{1,T}$ and $\vec{p}_{2,T}$ and 
we defined the following notations, 
\bea 
\langle \frac{1}{q^{2n}_T}\rangle&=&\frac{\int \frac{ d^2 \vec q_T}{q^{2n}_T}~\phi\Lb  Y-y,\vec q_T, - \vec q_T
\Rb \,\phi\Lb y,\vec{q}_T - \vec{p}_T, \vec{p}_T - \vec{q}_T\Rb} { \int  d^2 \vec q_T \phi\Lb  Y - y, \vec q_T, -\vec q_T
\Rb \,\phi\Lb y,\vec{q}_T - \vec{p}_T, \vec{p}_T - \vec{q}_T\Rb}, \label{2BFKLSH61}\\
\langle \langle Q^{2n}_T\rangle\rangle &=& \frac{\int d^2 \vec Q_T~ Q^{2n}_T \,N_{\pom h}^2(Q^2_T)} {\int d^2 \vec Q_T  \,N_{\pom h}^2(Q^2_T)},\label{2NBFKLSH62}
\eea
with $n=1,2$. The normalization factor  $\cal N$ in \eq{2BFKLSH5} is given by  
\beq \label{2BFKLSH6}
 {\cal N}\,\,\,\equiv\,\, \pi \int d Q^2_T \,N_{\pom h}^2(Q^2_T)\, \frac{d \sigma}{d y_1\,d^2 \vec p_{1,T}}\Lb Q_T = 0\Rb\,\frac{d \sigma}{d y_2\,d^2 \vec p_{2,T}}\Lb Q_T = 0\Rb. 
\eeq
From the above, it is obvious that the production of two
parton showers  with a transverse momentum $\vec Q_T$ along the Pomeron ladder,  naturally leads to the long range
rapidity correlation in azimuthal angle while the emissions from one parton
shower given by the BFKL Pomeron contribution does not lead to such correlations, see also Ref.~\cite{THEOREXP}. 

 \begin{figure}[t]
\begin{center}
       \includegraphics[width=10cm] {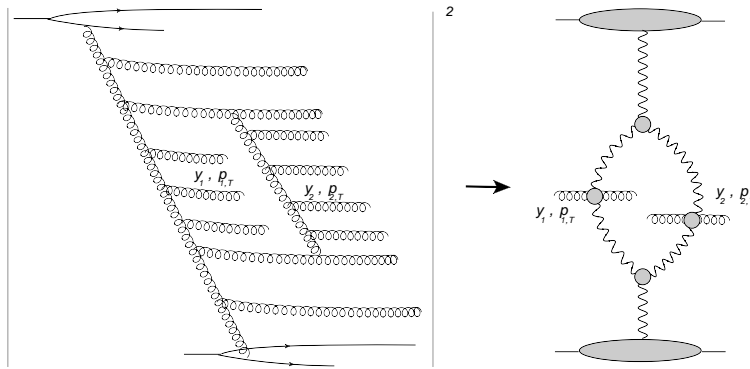}
\end{center}
\caption{Parton shower production with the typical enhanced diagram and the corresponding Mueller diagram for two gluon correlations. The wave lines denote the BFKL Pomerons.} 
\label{2pshen}
\end{figure}

\section{Estimates of azimuthal angle correlations in $pp$ and $AA$ collisions }
We recall that the long-range azimuthal angle correlations obtained by
\eq{2BFKLSH5} is valid in the leading log-s approximation at
high-energy. The azimuthal angle correlations in \eq{2BFKLSH5} is
uniquely determined by only knowing the average values $\langle
1/q^{2n}_T\rangle$ and $\langle\langle Q^{2n}_T\rangle
\rangle$. This equation was truncated at $n=2$ assuming that the transverse momentum $Q_T$ in the Pomeron ladder is small. Let us explore the idea that \eq{2BFKLSH5} is also
valid in the saturation region (or at least on the boundary between the BFKL
and the saturation regime) by choosing the corresponding average values
$\langle 1/q^{2n}_T\rangle$ and $\langle\langle Q^{2n}_T\rangle
\rangle$ in that region.

In the kinematic regime of the BFKL (ignoring the saturation effect) from
\eq{2BFKLSH61} we obtain $\langle 1/q^{2n}_{T}\rangle\,\approx
1/\text{max}\{\mu^{2n}, Q_T^{2n}\}$ where $\mu$ is the non-perturbative soft
scale.  At the LHC energies, the inclusive production stems
from the kinematic region in which saturation effects are important \cite{LRPP,LRAA1,LRAA2}.
In this region, the interaction between Pomerons leads to more
complicated diagrams, the so-called enhanced diagrams shown in
\fig{2pshen}. It has been shown (see Ref.~\cite{LMP} and references
therein) that the enhanced diagram leads to the value of the
characteristic momentum of the order of $Q_s$, namely we have $Q_T
\propto Q_s$ . We do not need to follow complete calculations of this
paper to understand why it happens so. Indeed, assuming that $Q_T
\,\ll\, \mbox{typical q $\approx Q_s$}$  we can replace
the BFKL Pomerons in the loop by the Pomerons at $Q_T=0$.  Therefore,
in this case, we have $\int^{Q_s} d^2 Q_T \,=\,Q^2_s$.  For $Q_T \gg q
$ the Pomeron exchange falls down with $Q_T$ making the integral being
concentrated at $Q_T = q_T = Q_s$. In order words, if densities of
partons in one parton shower is so large that we have already reached
the saturation region of the gluon density, we can assume that the
average $\langle 1/q^{2n}_T\rangle \approx 1/Q^{2n}_s$ where $Q_s$ is
the saturation scale. This also follows from the high-density QCD
within the CGC approach \cite{MV} which describes the LHC data for the
inclusive hadron production both in $pp$ and $AA$ collisions
\cite{LRPP,LRAA1}, see also Ref.~\cite{LRAA2}. We also assume that the
density of partons in {\it both} parton showers is very large and
consequently Pomeron enhanced diagrams are important, and therefore we
can have $\langle\langle Q^{2n}_T\rangle \rangle\approx Q^{2n}_s$.
Therefore, we assume that in the saturation region we have only one
relevant scale, the saturation scale, and the average transverse
momenta are related to this scale.

Notice that the maximum of the double inclusive production reaches at
$p_{1,T} \approx p_{2,T} \approx Q_s$.  Admittedly, we do not have a rigorous proof
of this at our disposal without invoking any approximation, but this
may be immediately understood within the CGC approach since $Q_s$ is
the only dimensional parameter of the approach. This can be also
seen in the simple case of the Born approximation by comparing
Eqs.~(\ref{BFKLEM3}, \ref{BFKLEM4}).  Note that \eq{BFKLEM3} gives the
contribution at small values of $p_T$ and the correlations vanish at
$p_t \to 0$ and increase with $p_T$ while \eq{BFKLEM4} shows that the
correlations falls down at large values of $p_T$. Therefore, the
correlation function has a maximum at $p_T \approx \langle q_T
\rangle$, where $\langle q_T \rangle$ is the typical transverse
momentum of the system. The same argument is valid for the general
case of the gluon pairs production from the BFKL Pomeron. This can be
seen by comparing \eq{2BFKLSH5} and its corresponding equation in the
limit of $p_T\gg q_T\gg Q_T$.  It should be stressed that the
experimental data from the CMS collaboration indicates that the
maximum of correlations occurs at the kinematic region that the
saturation effects is important
\cite{CMSLRC}.

The probability for the events with multiplicity equals $N = 2 \langle
N\rangle $ where $\langle N\rangle $ is the multiplicity in one parton
shower, can be obtained by \eq{2BFKLSH6} and the corresponding
cross-section of such events is $\sigma (N = 2 \langle N \rangle)
\propto {\cal N}$. Using \eq{2BFKLSH5}, we obtain two-particle
correlation function $\mathcal{R}$ for the event selections with
multiplicity $N$ as,
\beq \label{PP5}
\mathcal{R}\Lb \Delta \varphi; y_1,y_2\Rb\,\,=\,\, \frac{\frac{d N}{d y_1 d^2\vec p_{1,T} dy_2 d^2\vec p_{2,T}}}
{\frac{d^2 N}{d y_1 d^2\vec p_{1,T}} \, \frac{d^2 N}{d y_2 d^2\vec p_{2,T}}}\,\,\,-\,\,\,1
\,\,=\,\,\frac{\overline n(\overline n-1)}{2~\overline{n}^2}\,\,\Big\{ 1 \,\,+\,\,\frac{1}{2}\,\Lb 2 \,+\,\cos\left(2 \Delta\varphi\right)\Rb\Big\}\,\,-\,\,1, 
\eeq
where the parameter $\overline n=E\left(N/\langle N\rangle\right)$ is
the relative average number of Pomeron parton showers in the event selections
with multiplicity $N$, the average multiplicity $\langle N\rangle$
denotes the multiplicity in the mini-bias and function $E$ gives the
integer value of its argument. The pre-factor in \eq{PP5} comes from
the counting the various possible ways to have two gluons production
out of $\overline n$ Pomeron parton showers. In other words, for simplicity we
assumed that $\bar{n}$ showers are produced and two correlated gluons
comes from only two different parton showers. The number of these pairs is
equal to $\bar{n}\Lb \bar{n} - 1\Rb/2$ and moreover we have $\frac{d^2 N}{d y_i d^2 p_{i,T}} \,\,=\,\,\bar{n}
\frac{d^2 N\Lb \mbox{one parton shower}\Rb}{d y_i d^2 p_{i,T}}$, therefore the pre-factor in \eq{PP5} can be readily obtained.

Notice that the main background for the double inclusive gluons
production is due to two jets production from one parton shower. However,
this production is suppressed by making selection in the events. From
AGK cutting rules \cite{AGK}, it follows that the multiplicity in one
parton shower is equal to the average multiplicity measured by the
experiment in the mini-bias events. It should be stressed that the AGK
cutting rules also work for two parton showers production in QCD
\cite{AGKQCD7}. It is well-known that the gluon distribution in the
BFKL Pomeron is close to the Poisson distribution, see Ref.~\cite{LEN}
and references therein.  The production from two parton showers starts
to be significant only for the events with multiplicity larger than $2
\langle N \rangle$ where $\langle N \rangle$ is the mean multiplicity,
see Fig.~\ref{2psh}.  On the other hand, the probability to have 
events with multiplicity $ 2
\langle N \rangle$ in one parton shower is approximately suppressed as $ \exp\Lb - (
2 \langle N \rangle -
\langle N \rangle)^2//2 \langle N \rangle\Rb \,\ll\,1$ for the Poisson
distribution.

One can observe in \eq{PP5} that except the over-all pre-factor, the
coefficients does not depend on multiplicity and rapidity of pairs. Of
course this feature may be altered due to possible contamination of two
gluons production from one parton shower which may lead to short range
rapidity correlations in the azimuthal angle $\Delta\varphi$. However, in
particular experimental set up  with high multiplicity events where
our underlying saturation assumption namely $\langle \langle
Q^{2n}_T\rangle\rangle \approx \langle q^{2n}_T\rangle \approx Q_s^{2n}$ is at work, these
correlations could be ignored and can only create a back ground that
will fall off at large multiplicity events.

In order to understand how much the azimuthal asymmetry depends on the
value of $\langle \langle Q^{2n}_T\rangle\rangle$, we next
estimate $\langle \langle Q^{2n}_T\rangle\rangle$ in the BFKL
kinematic region ignoring the so-called enhanced diagrams (shown in
\fig{2pshen}). In order to calculate $\langle \langle
Q^{2n}_T\rangle\rangle$ defined in \eq{2NBFKLSH62}, we should know the
non-perturbative amplitude $N_{\pom p}\Lb Q_T\Rb$ defined in \eq{EQN}.  For
$N_{\pom p}\Lb Q_T\Rb$, we use the quasi-eikonal approximation
\cite{LRCORKT}. In this approximation we restrict ourselves to the
first term in \eq{EQN} and the contribution of the other terms is
taken into account by introducing an extra factor $N_0$, 
\beq \label{EIKN}
N_{\pom p}\Lb Q_T\Rb \,\,=\,\,N_0 g^2_{\pom p}\Lb Q_T\Rb,
\eeq
where $g_{\pom p}$ is the vertex of Pomeron-proton interaction. This
approximation has been widely employed in Pomeron phenomenology and
works quite well in the description of the experimental data
\cite{KMR}. The dependence of the BFKL Pomeron on the transverse momentum $Q_T$ is given by $g_{\pom p}\Lb Q_T\Rb=1/(1 + Q^2_T/m^2)^2$ with the typical mass $m$ determined from the experimental  data. 
The dipole form of $g_{\pom p}$ is inspired by the $Q_T$ dependence of the
electromagnetic form factor of the proton. Using this distribution
we obtain $\langle Q^2_T\rangle = m^2/6$ and $\langle Q^4_T\rangle =
m^2/15$. The experimental data for diffractive production of the vector meson in the DIS
\cite{HERAVP} indicates that $m^2 = 0.8 \,\text{GeV}^2$.  However, the CDF
data on double jet production \cite{CDFDPXS} shows that the 
typical value of $Q_T$ could be larger leading to a bigger value for $m^2 = 1.6
\,\text{GeV}^2$. Again assuming that $\langle q_T^{2n}\rangle=Q_s^{2n}$ the corresponding 
two-particle correlation function $\mathcal R$ for $p_{1T}=p_{2T}=Q_s$ and $\overline n\geq 2$ becomes, 
\beq \label{PP1}
\mathcal{R}\Lb \Delta\varphi; y_1,y_2\Rb\,\,
\,\,=\frac{\overline n(\overline n-1)}{2~\overline{n}^2}\,\,\Big\{\,\, 1 \,\,+\,\,\frac{m^4}{30\, Q^4_s}\,\Lb 2 \,+\,\cos\left( 2 \Delta\varphi\right)\Rb\,\,\Big\}-1.
\eeq
\begin{figure}[t]
        \includegraphics[width=17cm,height=10cm] {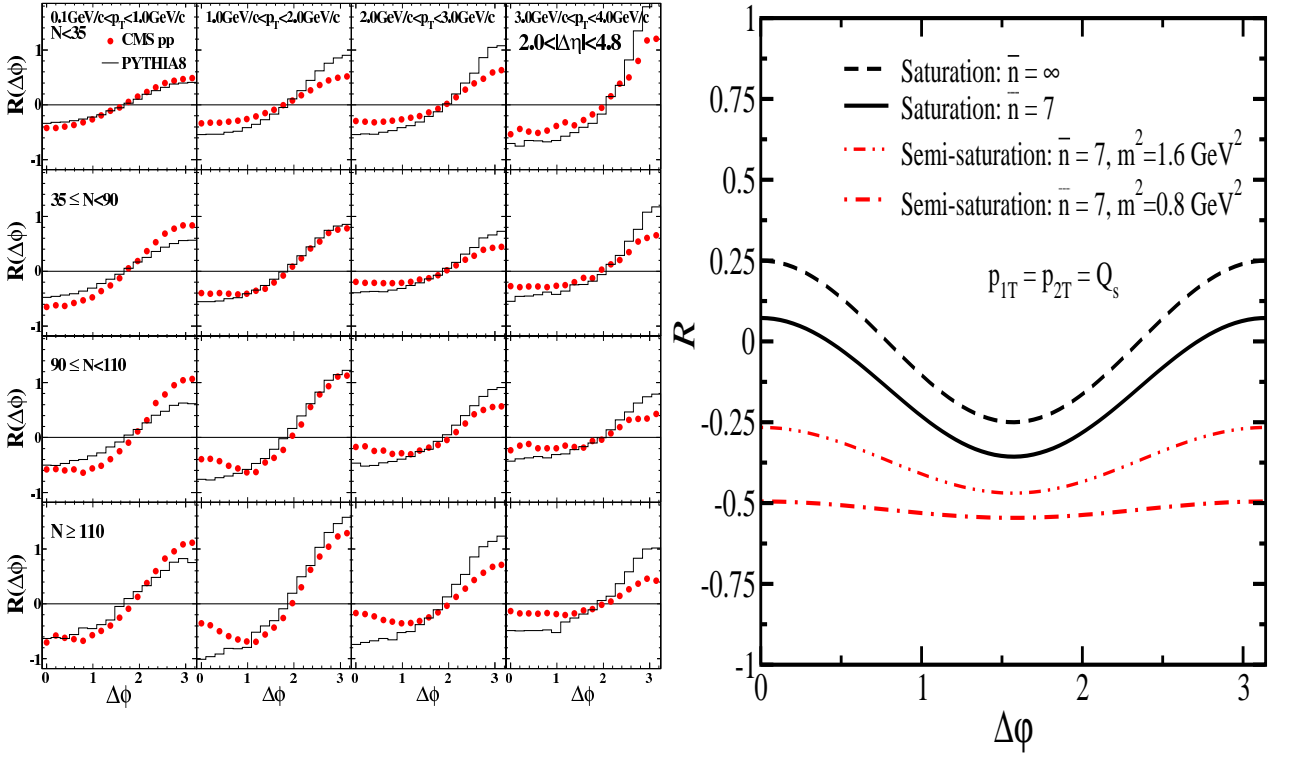}
\caption{Right: The correlation function $\mathcal{R}$ at different multiplicity $N=\overline n\langle N\rangle$. 
The curves labeled by ``Saturation'' are the results from \eq{PP5}
when $\langle \langle Q_T^{2n}\rangle\rangle\approx Q_s^{2n}$ with $Q_s$ being the
saturation scale. The curves labeled by ``Semi-saturation'' are the
results from \eq{PP1} when $\langle \langle Q_T^{2n}\rangle\rangle$ was
calculated within BFKL region for two different masses $m^2 =
0.8\,\text{GeV}^2$ and $m^2 = 1.6 \,\text{GeV}^2$. In both cases we
assumed $\langle 1/q^{2n}_T\rangle \approx 1/Q^{2n}_s$. Left:
Experimental data from the CMS collaboration for projections of 2-D
correlation functions onto $\Delta \phi$ for $2 < \Delta\eta < 4.8$ in
different $p_T$ and multiplicity bins at 7 TeV pp collisions and
reconstructed PYTHIA8 simulations \cite{pyt}.  Error bars are smaller
than the symbols. The plot in the left panel is taken from
Ref.~\cite{CMSLRC}.}
\label{ppcor}
\end{figure}
It is seen from above that the coefficients in $\mathcal R$ now
depends on the rapidity via the saturation scale $Q_s$ in contrast to
\eq{PP5}. However, one should note that deep inside the saturation region the above equation is not reliable 
and one should then use \eq{PP5}. It is instructive to notice that the
two BFKL parton showers contribution lead to \eq{PP1} with the soft
scale $\mu$ instead of $Q_s$. This scale is a new phenomenological
parameter which does not depend on energy, and it is certainly $\mu
\leq Q_s$.

In \fig{ppcor} (right) we show the azimuthal correlation
$\mathcal{R}$ obtained from
\eq{PP1} when $\langle \langle Q_T^{2n}\rangle\rangle$ was calculated within the BFKL region for two different
masses $m^2 = 0.8\,\text{GeV}^2$ and $m^2 = 1.6 \,\text{GeV}^2$. In
this plot, we take a fixed saturation scale $Q_s^2 = 0.6 \,\text{GeV}^2$. The
chosen saturation scale is in accordance with the estimates of
Ref.~\cite{LRPP} in $pp$ collisions at the LHC. In \fig{ppcor} (right) we also show the azimuthal
correlation $\mathcal{R}$ obtained from
\eq{PP5} in the saturation region at different multiplicity $N=\overline n\langle N\rangle$. 
It is observed that deep inside the saturation region we have the ridge-type
structure, namely a second local maximum near $\Delta \varphi\approx
0$ {\it independent of rapidity} when $p_T$ is about the saturation scale.  By comparing the results
shown in \fig{ppcor} from Eqs.~(\ref{PP5},\ref{PP1}), it is notably
seen that by going from the BFKL to the saturation region, a pronounced ridge-type
structure emerges.  In
\fig{ppcor} we show the experimental data from the CMS collaboration
\cite{CMSLRC} for projections of two dimensional correlation functions onto $\Delta \varphi$ (denoted in \fig{ppcor} left panel by $\Delta \phi$) for  the difference in pseudorapidity of pair $2 <
\Delta\eta < 4.8$ in different $p_T$ and multiplicity bins at 7 TeV pp
collisions and reconstructed PYTHIA8 simulations \cite{pyt}. It is
important to note that PYTHIA8 qualitatively fails to reproduce the
local maximum in the near-side correlation in any of the $p_T$ or multiplicity bins \cite{CMSLRC}, see \fig{ppcor}.
Notice that our definition of two-particle correlation $\mathcal{R}$
defined in Eq.~(\ref{PP5}) is different from the experimental
definition $R$
\cite{CMSLRC} shown in
\fig{ppcor} with a over-all factor.  Here given the simplicity of our approach we do not wish to compare directly our results with the
experimental data. A meaningful comparison requires inclusion of
correlations effect within one parton-shower, fragmentation and possible short-range
correlation effects. Nevertheless, it is seen that the general
feature of the near-side two-point correlations obtained by
Eqs.~(\ref{PP5},\ref{PP1}) is compatible with the CMS experimental
data \cite{CMSLRC}.  One should note that for a denser system, the
imposed condition of $p_{1T}=p_{2T}=Q_s$ in \eq{PP5} shifts the
relevant kinematic windows of the angular correlations to the higher
$p_T$ since the saturation scale will be larger for a denser system.

Next, we consider the long-range correlations in nucleus-nucleus
scatterings. It is straightforward to generalize \eq{2BFKLSH4} for the
case of nucleus-nucleus collisions in the framework of the Glauber
approach, namely assuming that multiple scatterings are only permitted
on different nucleons while the nucleon-nucleon scattering stems from
the BFKL Pomeron exchange. The double inclusive cross-section at a
fixed impact-parameter between two center of nuclei $b$ will then have
the same form as \eq{2BFKLSH4} except the extra dependence on the nuclear profile. Notice that the
impact-parameter $b$ is the conjugate variable to transverse momentum
$Q_T$. In the Glauber approximation for nuclei, the scattering
amplitude of Pomeron-nucleus $N_{\pom A}$ in the region of small diffractive masses is defined as
\bea \label{EQNA}
N_{\pom A}\Lb Q_T\Rb\,\,&\equiv & \left( \int \,d^2 \vec{b}~d^2\vec{b'} \,e^{i \vec{Q}_T \cdot \vec{b}}\,S_A\Lb \vec b-\vec{b'} \Rb g_{\pom p}(\vec{b'})\right)^2,\nonumber\\
&\approx&~ g^2_{\pom p}(q_T=0)~S^2_{A}(Q_T),\
\eea
where $g_{\pom p}$ denotes the Pomeron-proton vertex and $S_A\Lb b
\Rb$ is the nuclear density profile defined by the Wood-Saxon
parametrization. The second equation above is valid when the nuclear
radius is larger compared to the proton size $R_A>>R_p$. Using
\eq{EQNA} one can obtain the following expression for the double
inclusive cross-section at fixed $b$ in the framework of the Glauber
approach in which the proton-proton scatterings are taken into account
from the BFKL Pomeron:
\bea \label{AA1}
\frac{d \sigma_A}{ d y_1\,d y_2\,d^2 p_{1,T}\,d^2 p_{2,T}\,d^2 b}&=&\h\,\frac{d \sigma_N}{ d y_1\,d^2 p_{1,T} }\frac{d \sigma_N}{d y_2\,d^2 p_{2,T} }\nonumber\\
&\times &\Big\{ T_{AA}^2(b)
+\,\h\,p^2_{1,T}\,p^2_{2,T}\,\Big(\langle 1/q^4\rangle_{\text{proton}}\Big)^2\, \Big(\bigtriangledown^2_b\bigtriangledown^2_b\,T^2_{AA}(b) \Big)\Lb 2\,+\,\cos\left(2 \Delta\varphi\right)\Rb { \Big\}},\
\eea
where $\frac{d \sigma_N}{ d y_1\,d^2 p_{1,T} }$ is the inclusive
cross-section for proton-proton scatterings \footnote{It should be
noted that the inclusive cross-section of proton-proton scatterings
enters \eq{AA1} since the integration over the impact parameter of
proton-proton scatterings has been performed as usual in the Glauber
approach.} and $T_{AA}$ is the nuclear overlap function for AA collisions. 
Using \eq{AA1} one can calculate the correlation function $\mathcal R$ defined in \eq{PP5}:
\bea \label{AA3}
 \mathcal{R}\Lb b,\Delta\varphi, y_1,y_2\Rb \,\,&=&\,\,\,\h\,p^2_{1,T}\,p^2_{2,T}\,\Big(\langle 1/q^4\rangle_{\text{proton}}\Big)^2\, \Big(\bigtriangledown^2_b\bigtriangledown^2_b\,T^2_{AA}(b) \Big)\Lb 2\,+\,\cos\left(2 \Delta\varphi\right)\Rb. \
\eea
It is straightforward to show that in the Glauber approach, the inclusive
production in $AA$ collisions is proportional to the overlap function
$T_{AA}(b)$ while in the case of the double inclusive production
instead it is proportional to $T_{AA}^2(b)$. It is seen from \eq{AA3}
that independent productions are canceled in $\mathcal R$ at fixed
$b$ while in the integral over $b$ the first term in \eq{AA1} gives
the main contribution. It is worth mentioning that we do not need
additional factor $\overline n$ as in \eq{PP1} since in the Glauber
formulation for nucleus-nucleus scatterings, events with a fixed
multiplicity correspond to a definite value of the impact-parameter.

Deep inside the saturation region, the correlation function has the same form
as for hadron-hadron collisions given by \eq{PP5} with the saturation momentum
replaced by that of the nucleus $Q^2_s(AA;x)\,\,\approx
\,\,T_{AA}\Lb b \Rb\, Q^2_s\Lb pp;x\Rb$ \cite{LRAA1,LRAA2}. This is in agreement with the
main idea of the CGC approach that difference between different
reactions is only due to the different value of the saturation scale
$Q_s$.  Assuming that we have $\langle q^{2n} \rangle_{\text{proton}}=
Q_s^{2n}\Lb pp; x\Rb$ ($n=1,2$) in the saturation region, it is seen
from \eq{AA3} that two-particle correlation $\mathcal{R}$ reduces by increasing
the saturation scale of proton. However, the Glauber approximation
is not reliable deep inside the saturation region, and one should instead use
\eq{PP5}, consequently the slope of the reduction of the azimuthal correlations will be then different. One of the attractive feature of nucleus-nucleus 
collisions is that by using centrality cuts, one can study the
underlying dynamics of two-particle correlations.

\section{Conclusions}

In this paper, we suggested a new mechanism for the long-range
rapidity correlations in the azimuthal angle of produced hadron pairs,
namely the long-range angle correlations of two parton showers
component of the initial partonic (gluonic) wave-function. This
mechanism can be conceived as a realization of the general ideas proposed in 
Refs. \cite{KOVLUCOR,NEWCORPP}.  Our approach predicts large and of
the same order long-range angular correlations both for hadron-hadron
and nucleus-nucleus collisions inside the gluon saturation region. In
our approach, the collimation in $\Delta \varphi$ exists independently
of the effects from flow in the later stages of the collisions. We
showed that for extremely dense systems at the truncation level upto $n=2$ for $\langle\langle Q^{2n}_T\rangle
\rangle$ we have $\mathcal{R}\to
0.25$ ($\mathcal{R}\to -0.5$ without correlation) at $\Delta\varphi
\approx 0$, and $\mathcal{R}$ still has a second local maximum
near $\Delta \varphi\approx 0$ at $p_T\approx Q_s$. We showed that
our mechanism qualitatively describes the main features of the
observed ridge structure in proton-proton collisions at the LHC at
$\sqrt{s}=7$ TeV. A detailed comparison with experimental data and
numerical analysis is left for future.

The main difference between our approach and the description in the
framework of the CGC \cite{THEOREXP,more-r,NEWCOR1,NEWCORPP}
is that in our approach the saturation region is explored from outside
on the boundary with the BFKL region. We showed that a clear signal of the ridge-type structure
emerges by going from the BFKL to the saturation regime. 
 This is fully consistent with the fact that the saturation/CGC
approach provides an adequate description of other $7$ TeV data
in $pp$ collisions including the inclusive charged-hadron
transverse-momentum and multiplicity distribution
\cite{LRPP,LRAA1}. Finally notice that the correlations obtained in
our approach is not suppressed with $1/N_c$ in contrast to the
prescription of Refs.~\cite{NEWCOR1,NEWCORPP} and survive in the
leading order in $1/N_c$ expansion.

 \section*{Acknowledgments}

 A. R. would like to thank Alex Kovner and Michael Lublinsky for useful
 discussion and remarks. This work was supported in part by the
 Fondecyt (Chile) grants 1100648 and 1110781.




\begin{thebibliography}{99}
\bibitem{CMSnew}
CMS Collaboration, arXiv:1105.2438. 

\bibitem{CMSLRC}
 V.~Khachatryan {\it et al.}  [CMS Collaboration],
  JHEP {\bf 1009} (2010) 091
  [arXiv:1009.4122].

\bibitem{NEWCORPP}
  A.~Dumitru, K.~Dusling, F.~Gelis, J. Jalilian-Marian, T. Lappi and R. Venugopalan 
Phys. Lett. {\bf B697} (2011) 21.

\bibitem{ppm}
K. Werner, Iu. Karpenko, T. Pierog, Phys. Rev. Lett. {\bf 106} (2011) 122004; P. Bozek, arXiv:1010.0405. 

\bibitem{rest}
I. O. Cherednikov and N. G. Stefanis, arXiv:1010.4463; Igor M. Dremin,
Victor T. Kim, arXiv:1010.0918; S. M. Troshin and N. E. Tyurin,
arXiv:1009.5229; I. Bautista, J. Dias de Deus and C. Pajares, e-Print:
arXiv:1011.1870; M. Y. Azarkin, I. M. Dremin and A. V. Leonidov, arXiv:1102.3258;
M. Diehl and A. Schafer, Phys. Lett. {\bf B698} (2011) 389;
J. Bartels and M. G. Ryskin, arXiv:1105.1638; 
S. Vogel, P. B. Gossiaux, K. Werner and J. Aichelin, arXiv:1012.0764;
R. C. Hwa and C. B. Yang, Phys. Rev. {\bf C83} (2011) 024911;
E. Avsar, C. Flensburg, Y. Hatta, J-Y Ollitrault and T. Ueda,
arXiv:1009.5643; H. R. Grigoryan and Y. V. Kovchegov, JHEP {\bf 1104} (2011) 010. 


\bibitem{RHICCOR}
J. Adams {\it et al.} [STAR Collaboration], Phys. Rev. Lett.
{\bf 95} (2005) 152301; J. Adams, {\it et al.} [STAR Collaboration], Phys. Rev. {\bf C73} (2006) 064907; A. Adare {\it et al.} [PHENIX Collaboration], Phys. Rev. {\bf C78} (2008) 014901; B. I. Abelev {\it et al.} [STAR Collaboration], Phys. Rev. {\bf C80} (2009) 064912;
B. Alver {\it et al.} [PHOBOS Collaboration], Phys. Rev. Lett. {\bf 104} (2010) 062301.



\bibitem{THEOREXP}
  Y.~V.~Kovchegov, E.~Levin and L.~D.~McLerran,
  Phys.\ Rev.\  {\bf C63}  024903 (2001)
  [hep-ph/9912367];\,\,\,
 D.~Kharzeev, E.~Levin and L.~McLerran,
  Nucl.\ Phys.\  {\bf A748} (2005) 627
  [hep-ph/0403271];\,\,\,
N.~Armesto, L.~McLerran and C.~Pajares,
  Nucl.\ Phys.\   {\bf A781} (2007) 201
  [hep-ph/0607345];\,\,\,
K.~Fukushima and Y.~Hidaka,
  Nucl.\ Phys.\   {\bf A813} (2008) 171
  [arXiv:0806.2143].






\bibitem{NEWCOR1}
A. Dumitru, F. Gelis , L. McLerran and R. Venugopalan, Nucl. Phys. {\bf A810} (2008) 91 [arXiv:0804.3858];
K. Dusling, F. Gelis, T. Lappi and R. Venugopalan, Nucl. Phys. {\bf A836} (2010)159 [arXiv:0911.2720]. 




\bibitem{more-r}
A. Dumitru and J. Jalilian-Marian, Phys. Rev. {\bf D81} (2010) 094015 (2010);
 T.~Lappi and L.~McLerran,
 Nucl.\ Phys.\  {\bf A832 } (2010) 330 [arXiv:0909.0428];\,\,\, F.~Gelis, T.~Lappi and L.~McLerran,
  Nucl.\ Phys.\   {\bf A828} (2009) 149
  [arXiv:0905.3234];
G.~Moschelli, S.~Gavin and L.~McLerran,
  Eur.\ Phys.\ J.\   {\bf C62} (2009) 277;
J. Jalilian-Marian,  arXiv:1011.1601; T. Lappi,  arXiv:1011.0821. 






\bibitem{MV}
L. McLerran and R. Venugopalan, Phys. Rev. {\bf D49} (1994) 2233; {\bf
D49} (1994) 3352; {\bf D50}, (1994) 2225; {\bf D53} (1996) 458; {\bf
D59} (1999) 094002. For a recent review see: R. Venugopalan,
arXiv:1012.4699; L. McLerran, arXiv:1011.3203; F. Gelis, E. Iancu,
J. Jalilian- Marian and R. Venugopalan, arXiv:1002.0333. 


\bibitem{dav}
D. d'Enterria, G. Kh. Eyyubova, V. L. Korotkikh, I. P. Lokhtin,
S. V. Petrushanko, L. I. Sarycheva, A. M. Snigirev, Eur. Phys. J. {\bf C66} (2010) 173 
[arXiv:0910.3029]. 
\bibitem{mev2}
B. Z. Kopeliovich, A. H. Rezaeian and I. Schmidt, Phys. Rev. {\bf D78}
(2008) 114009 [arXiv:0809.4327].


\bibitem{KOVLUCOR}
 A.~Kovner, M.~Lublinsky, Phys. Rev. {\bf D83} (2011) 034017 
  [arXiv:1012.3398].




\bibitem{BFKL}
 E. A. Kuraev, L. N. Lipatov, and F. S. Fadin,  Sov. Phys.
JETP {\bf 45} (1977) 199; Ya. Ya. Balitsky and L. N. Lipatov,
Sov. J. Nucl. Phys.\, {\bf 28} (1978) 22.

\bibitem{GLR}
L. V. Gribov, E. M. Levin and M. G. Ryskin, Phys. Rep.\,
{\bf 100} (1983) 1.

\bibitem{MUQI}
A. H. Mueller and J. Qiu,  Nucl. Phys.  {\bf B268}
(1986) 427.

\bibitem{BRN}
M.~A.~Braun, Phys.\ Lett.\, {\bf B632} (2006) 297
  [hep-ph/0512057];\,\,
arXiv:hep-ph/0504002\,;
Eur.\ Phys.\ J.  {\bf C16} (2000) 337 
[hep-ph/0001268];\,\,\,
  Phys.\ Lett.\  {\bf B483} (2000) 115
  [hep-ph/0003004];\,\,
  Eur.\ Phys.\ J.\  {\bf C33} (2004) 113
  [hep-ph/0309293];\,\,\,
Eur.\ Phys.\ J.  {\bf C6}  (1999) 321
[hep-ph/9706373];\,\,\,
M.~A.~Braun and G.~P.~Vacca,
Eur.\ Phys.\ J.  {\bf C6} (1999) 147 [hep-ph/9711486].

\bibitem{BART}
J.~Bartels, M.~Braun and G.~P.~Vacca,
 Eur.\ Phys.\ J.  {\bf C40} (2005) 419 
  [arXiv:hep-ph/0412218];
J.~Bartels and C.~Ewerz,
JHEP {\bf 9909}, 026 (1999)
[hep-ph/9908454]; 
J.~Bartels and M.~Wusthoff,
 Z.\ Phys. {\bf C66} (1995) 157, 
A.~H.~Mueller and B.~Patel, Nucl.\ Phys.  {\bf B425} (1994) 471 
[hep-ph/9403256];
J.~Bartels, Z.\ Phys.\  {\bf C60} (1993) 471.
\bibitem{LMP}
  E.~Levin, J.~Miller and A.~Prygarin,
  Nucl.\ Phys.\  {\bf A806} (2008) 245.

\bibitem{MUDIA}
A. H. Mueller, Phys. Rev. {\bf D2} (1970) 2963.

\bibitem{COL}
P. D. B. Collins, {\it "An introduction to Regge theory and high energy physics"},
Cambridge University Press 1977.
\bibitem{SOFT}
L. Caneschi (editor), {\it "Regge Theory of Low -$p_T$ Hadronic Interaction"},
North-Holland 1989.
\bibitem{LRCORLR}
 E.~M.~Levin, M.~G.~Ryskin and N.~N.~Nikolaev,
  Z.\ Phys.\  {\bf C5} (1980) 285;\,\,\,
  E.~M.~Levin, M.~G.~Ryskin and S.~I.~Troian,
  Sov.\ J.\ Nucl.\ Phys.\  {\bf 23} (1976) 222
  [Yad.\ Fiz.\  {\bf 23} (1976) 423];\,\,\,
 E.~M.~Levin and M.~G.~Ryskin,
  Sov.\ J.\ Nucl.\ Phys.\  {\bf 20} (1975) 280
  [Yad.\ Fiz.\  {\bf 20} (1974) 519];
\,\,\,
E.~L.~Berger and M.~Jacob,
  Phys.\ Rev.\  {\bf D6} (1972) 1930.

\bibitem{LRCORKT}
 A.~B.~Kaidalov and K.~A.~Ter-Martirosian,
  Sov.\ J.\ Nucl.\ Phys.\  {\bf 39} (1984) 979
  [Yad.\ Fiz.\  {\bf 39} (1984) 1545];\,\,\,
A.~B.~Kaidalov,
  Phys.\ Rept.\  {\bf 50} (1979) 157.
\bibitem{LRCORORSAY}
 P.~Aurenche, F.~W.~Bopp, A.~Capella, J.~Kwiecinski, M.~Maire, J.~Ranft and J.~Tran Thanh Van,
  Phys.\ Rev.\  {\bf D45} (1992) 92;\,\,\,
  F.~W.~Bopp, A.~Capella, J.~Ranft and J.~Tran Thanh Van,
  Z.\ Phys.\  {\bf C51} (1991) 99;\,\,\,
  A.~Capella, C.~Pajares and A.~V.~Ramallo,
  Nucl.\ Phys.\  {\bf B241} (1984) 75;\,\,\,
  A.~Capella and J.~Tran Thanh Van,
  Phys.\ Rev.\   {\bf D29} (1984) 2512;\,\,\.
A.~Capella and J.~Tran Thanh Van
  Z.\ Phys.\  {\bf C18} (1983) 85;\,\,\,\,
 A.~Capella and A.~Krzywicki,
  Phys.\ Rev.\  {\bf D18} (1978) 4120;\,\,\,
A.~Capella, U.~Sukhatme, C.~I.~Tan and J.~Tran Thanh Van,
  Phys.\ Rept.\  {\bf 236} (1994) 225.

\bibitem{AGK}
 V.~A.~Abramovsky, V.~N.~Gribov, O.~V.~Kancheli, Yad.\ Fiz.\ {\bf 18 } (1973) 595 
 [Can be found in Ref.~\cite{SOFT}].

\bibitem{yura}

Y. V. Kovchegov and K. Tuchin,Phys. Rev. {\bf D65} (2002) 074026.
Y. V. Kovchegov, Phys. Rev. {\bf D72} (2005) 094009. 


\bibitem{LR2REV}
 E.~M.~Levin and M.~G.~Ryskin,
  Phys.\ Rept.\  {\bf 189 } (1990) 267 ; L.~N.~Lipatov,
  Phys.\ Rept.\  {\bf 286 } (1997) 131 and references therein.




\bibitem{LRPP}
E.~Levin and  A. H. Rezaeian,
  Phys.\ Rev.\  {\bf D82 } (2010)  014022 [arXiv:1005.0631]; see also: arXiv:1011.3591.
\bibitem{LRAA1}
E. Levin and A. H. Rezaeian, Phys. Rev. {\bf D83} (2011) 114001 [arXiv:1102.2385].  
\bibitem{LRAA2}
E. Levin and A. H. Rezaeian, Phys. Rev. {\bf D82} (2010) 054003 [arXiv:1007.2430]. 


\bibitem{KMR}
 M.~G.~Ryskin, A.~D.~Martin, V.~A.~Khoze {\it et al.},
  J.\ Phys.\ {\bf G36 } (2009)  093001
  [arXiv:0907.1374];\,\,\,
  Eur.\ Phys.\ J.\  {\bf C60 } (2009) 265
  [arXiv:0812.2413];\,\,\,
  Eur.\ Phys.\ J.\  {\bf C60 } (2009) 249
  [arXiv:0812.2407];\,\,\,
Eur. Phys. J. {\bf C54} (2008) 199 [arXiv:0710.2494] and references therein. 

\bibitem{HERAVP}
H.~Kowalski and D.~Teaney,   
Phys.\ Rev.\ {\bf D68 } (2003) 114005 and references therein.

\bibitem{CDFDPXS}
  F.~Abe {\it et al.} [ CDF Collaboration ],
  Phys.\ Rev.\  {\bf D56 } (1997)  3811-3832.



\bibitem{AGKQCD7}
 E.~Levin and A.~Prygarin,
  Phys.\ Rev.\  {\bf C78} (2008) 065202
  [arXiv:0804.4747].
\bibitem{LEN}
 E.~Levin,
  Phys.\ Rev.\  {\bf D49 } (1994) 4469. 
\bibitem{pyt}
T. Sjostrand, S. Mrenna, and P. Z. Skands, Comput.
Phys. Commun. {\bf 178} (2008) 852 [arXiv:0710.3820]; see also {\it "PYTHIA 8 status"} by T. Sjostrand in: Alessandro {\it et al.}, arXiv:1101.1852. 

\end{thebibliography}
\end{document}